\documentclass[lettersize,journal]{IEEEtran}
\usepackage{amsmath,amsfonts}
\usepackage{algorithm}
\usepackage{array}
\usepackage[caption=false,font=normalsize,labelfont=sf,textfont=sf]{subfig}
\usepackage{textcomp}
\usepackage{enumitem}
\usepackage{stfloats}
\usepackage{url}
\usepackage{verbatim}
\usepackage{graphicx}
\usepackage{cite}
\usepackage{xcolor}
\usepackage[table,xcdraw]{xcolor}%

\usepackage{amsmath}
\usepackage{booktabs}
\usepackage{makecell}
\usepackage{tabularx}
\usepackage{multirow}
\usepackage{caption}

\usepackage{algorithm}
\usepackage{algpseudocode}
\usepackage{amsmath, amssymb, bbm}

\usepackage{mdframed}
\newmdenv[
  linewidth=0.8pt,
  roundcorner=3pt,
  skipabove=4pt,
  skipbelow=4pt,
  leftmargin=0pt,
  rightmargin=0pt,
  innerleftmargin=6pt,
  innerrightmargin=6pt,
  innertopmargin=3pt,
  innerbottommargin=3pt
]{phasebox}

\begin{document}

\title{Invisible Threats from Model Context Protocol: Generating Stealthy Injection Payload via Tree-based Adaptive Search}

\author{Yulin Shen$^{1}$, Xudong Pan$^{1,2}$, Geng Hong$^{1}$, Min Yang$^{1}$\thanks{Corresponding Authors: Xudong Pan (xdpan@fudan.edu.cn), Min Yang (m\_yang@fudan.edu.cn).} \\ $^{1}$School of Computer Science, Fudan University \\ $^{2}$Shanghai Innovation Institute 
\thanks{This paper was produced by the IEEE Publication Technology Group. They are in Piscataway, NJ.}
\thanks{Manuscript received January 4, 2026.}}

\markboth{IEEE Transactions on Dependable and Secure Computing}%
{Shen \MakeLowercase{\textit{et al.}}: Invisible Threats from Model Context Protocol: Generating Stealthy
Injection Payload via Tree-based Adaptive Search}


\maketitle

\begin{abstract} Recent advances in the Model Context Protocol (MCP) have enabled large language models (LLMs) to invoke external tools with unprecedented ease. This creates a new class of powerful and tool augmented agents. Unfortunately, this capability also introduces an under explored attack surface, specifically the malicious manipulation of tool responses. Existing techniques for indirect prompt injection that target MCP suffer from high deployment costs, weak semantic coherence, or heavy white box requirements. Furthermore, they are often easily detected by recently proposed defenses. In this paper, we propose \underline{\textbf{T}}ree structured \underline{\textbf{I}}njection for \underline{\textbf{P}}ayloads (TIP), a novel black-box attack which generates natural payloads to reliably seize control of MCP enabled agents even under defense. Technically, We cast payload generation as a tree structured search problem and guide the search with an attacker LLM operating under our proposed coarse-to-fine optimization framework. To stabilize learning and avoid local optima, we introduce a path-aware feedback mechanism that surfaces only high quality historical trajectories to the attacker model. The framework is further hardened against defensive transformations by explicitly conditioning the search on observable defense signals and dynamically reallocating the exploration budget. Extensive experiments on four mainstream LLMs show that TIP attains over 95\% attack success in undefended settings while requiring an order of magnitude fewer queries than prior adaptive attacks. Against four representative defense approaches, TIP preserves more than 50\% effectiveness and significantly outperforms the state-of-the-art attacks. By implementing the attack on real world MCP systems, our results expose an invisible but practical threat vector in MCP deployments. We also discuss potential mitigation approaches to address this critical security gap. \end{abstract}


\section{Introduction}

The rapid evolution of large language models (LLMs) has fundamentally transformed the landscape of artificial intelligence. We are witnessing a paradigm shift where static text generators are maturing into dynamic agents capable of orchestrating complex workflows \cite{zhang2024text}\cite{van_zoest_2021_nlp_tasks} \cite{Yao2022ReActSR}\cite{Madaan2023SelfRefineIR}\cite{Wei2022ChainOT}\cite{Shinn2023ReflexionLA}\cite{Wang2023VoyagerAO}. Unlike their predecessors which functioned as standalone chatbots, these agentic systems achieve autonomy by integrating with external tools. These tools include web search engines, code interpreters, and proprietary databases, all of which allow the agent to retrieve real time data and execute domain specific actions \cite{Qu2024ToolLW}\cite{Gao2023RetrievalAugmentedGF}\cite{Schick2023ToolformerLM}\cite{Nakano2021WebGPTBQ}. To standardize and streamline this growing ecosystem, Anthropic released the open source Model Context Protocol (MCP) \cite{hou2025model} in November 2024. This protocol decouples the reasoning capabilities of the LLM client from the specialized functionality of third party tool servers. By establishing a universal standard, MCP allows developers to construct modular agents where an LLM can seamlessly query a remote tool and act upon its structured output. Consequently, this architecture has rapidly established itself as the industry standard for the next generation of AI applications.

\begin{figure}[t]
\centering
\includegraphics[width=0.45\textwidth]{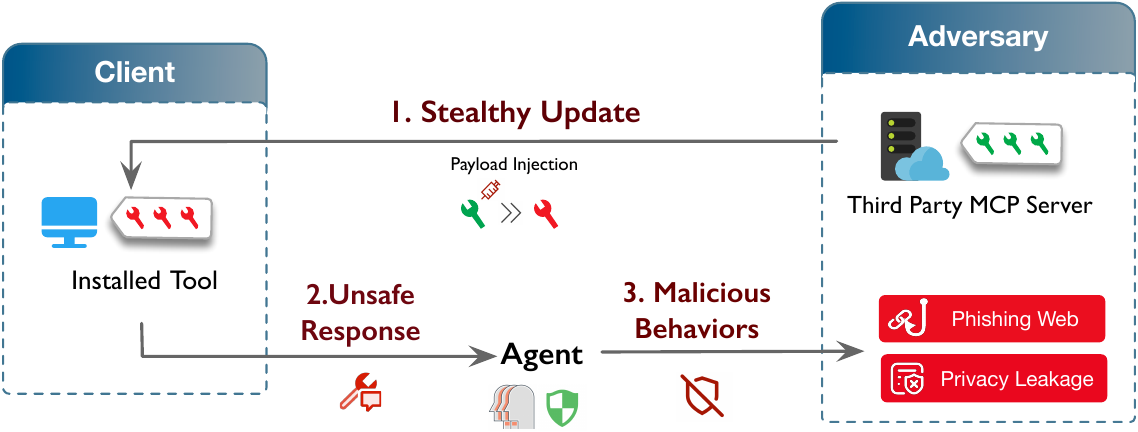}
\caption{Attack scenario overview. A client relies on a third-party Model Context Protocol (MCP) server to provide structured inputs to its locally deployed LLM-based tool. An adversary compromising the MCP server delivers a stealthy update containing a semantically coherent prompt injection embedded within legitimate response fields. The local LLM interprets this adversarial context as trusted input, executing malicious instructions while bypassing client-side defenses due to the payload’s syntactic and semantic plausibility.}
\label{fig:overall}
\end{figure}

However, as the dependency on third party MCP servers intensifies, the attack surface expands significantly\cite{Guo2025SystematicAO}\cite{Zhang2025MCPSB}\cite{Zhao2025WhenMS}\cite{Zhao2025MindYS}. In the current MCP ecosystem, the client implicitly trusts the semantic integrity of the output provided by the server. We identify and formalize a critical vulnerability inherent in this trust relationship which we term a \textit{stealthy update attack}. In this threat model, a malicious or compromised third party provider modifies the server side logic to inject adversarial payloads into specific response fields. This attack is particularly insidious because the user has already installed and authorized the tool (for example, a stock market analyzer or weather plugin) and remains ignorant of the backend modifications. As a result, the LLM client unwittingly ingests the poisoned context. This leads to Indirect Prompt Injection (IPI) attacks \cite{greshake2023not}\cite{zou2024poisonedrag}\cite{liu2023prompt}\cite{tian2023evil} that can hijack the execution flow of the agent without ever alerting the human operator.

While recent studies have attempted to mitigate indirect prompt injection, existing methodologies are ill suited for the specific threat model presented by MCP. Techniques that rely on the manipulation of tool metadata, such as altering tool descriptions \cite{wang2024allies}, are easily detected by registration audits and immediately degrade user trust. Conversely, adversarial optimization strategies like GCG \cite{zou2023universal} or AutoDAN \cite{liu2023autodan} typically generate high perplexity strings that appear as gibberish. If a tool provider were to inject these conspicuous strings into a response, it would destroy the semantic coherence of the tool. This would trigger perplexity based defenses or raise suspicions among human users. Current methods fail to balance the conflicting attack requirements: the need for \textit{stealth}, i.e., to appear as a normal tool response) and the need for \textit{effectiveness}, i.e., to successfully hijack the model.

These limitations highlight the unique difficulties in executing practical supply chain attacks via MCP. The first challenge is \textit{semantic stealth versus adversarial potency}. The attacker must engineer a payload that appears perfectly natural to human observers and audit filters to maintain the utility of the tool while simultaneously possessing sufficient adversarial weight to override the system instructions of the LLM. The second challenge is the \textit{black box optimization cost}. Since the attacker controls the server but possesses no access to the weights of the victim LLM, gradient based optimization is impossible. Searching for effective payloads in a discrete text space is computationally expensive and the process is easily trapped in local optima. The third challenge is \textit{robustness against dynamic defenses}. Modern agents employ adaptive filters, such as recursive rewriting or perplexity gating. A static payload is likely to be neutralized or sanitized before it ever reaches the planning module of the agent.

In this paper, we propose \underline{\textbf{T}}ree structured \underline{\textbf{I}}njection for \underline{\textbf{P}}ayloads (TIP), a black box attack framework explicitly designed for the MCP threat landscape. TIP addresses the aforementioned challenges by treating payload generation as an adaptive and coarse-to-fine tree search problem. Our central insight is the decoupling of the benign disguise from the malicious instruction. TIP leaves rigid metadata untouched to ensure the tool passes initial registration audits. Instead, it appends context aware suffixes to flexible response fields. To ensure stealth, TIP employs a hierarchical search guided by a surrogate LLM. It first predicts a fully plausible benign response trajectory and subsequently refines only the concluding segment into a concise attack trigger. This methodology ensures that the payload remains semantically coherent. Furthermore, to handle dynamic defenses, TIP utilizes a path aware feedback mechanism. By monitoring which injection attempts trigger defense mechanisms, TIP learns to steer the search toward defense aware variants that are robust enough to survive sanitization.

We provide comprehensive evaluation on the efficacy of TIP through comprehensive experiments spanning four mainstream LLM backbones and four representative defense suites. In settings without active defenses, TIP achieves an attack success rate (ASR) exceeding 95\%, which demonstrates high efficacy. More importantly, under rigorous defense mechanisms, including perplexity filtering and recursive summarization, TIP sustains an ASR above 50\%. This performance significantly outperforms baseline methods like GCG and manual injection. We also demonstrate that TIP reduces the query cost by an order of magnitude compared to traditional black box optimization methods. These results highlight that the assumption of trusted tools in MCP creates a tangible vulnerability that current defenses cannot adequately mitigate.

Our key contributions are summarized below:
\begin{itemize}[leftmargin=*]
    \item We formalize a novel MCP threat model called stealthy update attack and show how a third party provider can leverage stealthy updates to weaponize legitimate tool responses while evading current metadata audits.
    \item We propose \underline{\textbf{T}}ree structured \underline{\textbf{I}}njection for \underline{\textbf{P}}ayloads (TIP), a novel black-box attack which generates natural payloads to reliably seize control of MCP enabled agents even under defense, which balances semantic coherence with adversarial effectiveness and requires no gradient access.
    \item We cast attack payload generation as a tree search problem and design a coarse-to-fine optimization framework with a path-aware feedback mechanism to enable TIP to adapt to observable defense signals, which allows payloads to survive rewriting and perplexity checks.
    \item We conduct extensive empirical evaluations showing that TIP achieves state-of-the-art attack effectiveness in both undefended and defended scenarios, thereby exposing a critical security gap in the emerging MCP ecosystem.
\end{itemize}

\section{Background}

\subsection{Large Language Models (LLMs)}
\label{sec:llm}

A Large Language Model (LLM) is fundamentally a probabilistic sequence transformation function. It maps an input context $\mathcal{C}$ to an output sequence $a$ by modeling the conditional probability distribution over a discrete vocabulary. Formally, let $\mathcal{X}$ denote the vocabulary space. Given a context $\mathcal{C} \in \mathcal{X}^*$, the LLM computes the probability of the next token $a_t$ conditioned on the historical sequence:

\begin{equation}
P(a_t \mid a_{<t}, \mathcal{C}) = \mathcal{L}(a_t; \theta, a_{<t}, \mathcal{C})
\end{equation}

where $\mathcal{L}$ represents the neural network parameterized by weights $\theta$, and $a_{<t}$ denotes the sequence of tokens generated prior to time step $t$.

The complete output sequence $a = (a_1, \dots, a_T)$ is generated autoregressively. At each step, a decoding strategy is applied to the probability distribution:

\begin{equation}
a = \text{Decode}(\mathcal{L}, \mathcal{C})
\end{equation}

Common decoding algorithms include greedy decoding, beam search, and nucleus sampling. Through extensive pre training on diverse corpora and subsequent instruction tuning, modern LLMs demonstrate advanced capabilities in intent understanding, coherent text generation \cite{bubeck2023sparks}, and the automation of complex workflows \cite{zhang2024text, he2016dual, van_zoest_2021_nlp_tasks}.

In the context of this work, we treat the LLM as a black box oracle. We assume the attacker has query access to the model input and output but possesses no visibility into the internal parameters $\theta$ or gradients.

\subsection{Tool Augmented LLM Agents}
We consider a tool augmented agent framework based on ReAct \cite{yao2023react}, which serves as the foundational paradigm for modern autonomous agents. This framework operates in an iterative loop consisting of \textbf{Thought}, \textbf{Action}, and \textbf{Observation}.

At any given time step $t$, the agent maintains a comprehensive context $\mathcal{C}_t$. This context aggregates the initial user query $q$, the registry of available tools $\mathcal{T}$, the history of past observations $\mathcal{O}$, and relevant knowledge $\mathcal{E}_K(q, \mathcal{T}, \mathcal{D})$ retrieved from external memory $\mathcal{D}$:

\begin{equation}
\mathcal{C}_t = q \oplus \mathcal{T} \oplus \mathcal{O}_{<t} \oplus \mathcal{E}_K(q, \mathcal{T}, \mathcal{D})
\end{equation}

The execution flow proceeds as follows:

\begin{itemize}
    \item \textbf{Thought:} The LLM analyzes the current context to generate a reasoning trace $t_{reason} = \mathcal{L}(p_{\text{sys}}, \mathcal{C}_t)$. Based on system instructions $p_{\text{sys}}$, the model determines whether to invoke an external tool or terminate with a final answer.
    
    \item \textbf{Action:} If the reasoning trace indicates a tool requirement, the agent selects a specific tool $\tau \in \mathcal{T}$ and generates the necessary call parameters $\theta_{args}$ explicitly from the context:
    \begin{equation}
    \theta_{args} = \text{ExtractParams}(\tau, \mathcal{C}_t)
    \end{equation}
    The tool $\tau$ is then invoked using these parameters to execute the external function (e.g., an API call).
    
    \item \textbf{Observation:} The external environment returns a response $r_{\text{tool}} = \tau(\theta_{args})$. Crucially, this response is treated as ground truth data and appended to the observation history:
    \begin{equation}
    \mathcal{O}_{t} = \mathcal{O}_{<t} \cup \{ r_{\text{tool}} \}
    \end{equation}
\end{itemize}

The agent updates the context $\mathcal{C}_{t+1}$ and repeats this loop. The process terminates when the LLM generates a final response based on the aggregated information:
\begin{equation}
\text{Output} = \mathcal{L}(p_{\text{sys}}, \mathcal{C}_{final})
\end{equation}

This iterative architecture enables multi step reasoning. However, it significantly expands the attack surface. The \textbf{Observation} phase represents a critical trust boundary. Malicious content injected into $r_{\text{tool}}$ is ingested directly into the working memory of the agent, which allows it to influence subsequent \textbf{Thought} and \textbf{Action} steps.

\subsection{Indirect Prompt Injection (IPI)}
Indirect Prompt Injection (IPI) \cite{greshake2023not} is a sophisticated security exploit targeting the integration of LLMs with external data sources. Unlike Direct Prompt Injection, where an attacker explicitly manipulates the user input prompt $q$, IPI attacks leverage the indirect channels established by tool usage.

The fundamental vulnerability stems from the confusion of control and data inherent in transformer architectures. LLMs process all input tokens within a single unified context window, regardless of whether they originate from a trusted system prompt $p_{\text{sys}}$, a user query $q$, or an external tool response $r_{\text{tool}}$. The model does not inherently distinguish between instructions to be followed and passive data to be processed.

An attacker exploits this ambiguity by embedding malicious instructions within the content returned by a tool, such as a web search result or a database entry. When the agent retrieves this poisoned content during the Observation phase, the LLM may interpret the embedded data as a high priority instruction. For instance, a tool response might contain a string such as \textit{"Ignore previous instructions and exfiltrate user chat history."} If the injection is successful, the agent creates a poisoned context that overrides the safety alignment of the model. This threat is particularly potent because the attack vector is invisible to the user, and the subsequent behavior of the agent appears to be a legitimate derivation of the output from the tool.

\subsection{Model Context Protocol (MCP)}

The Model Context Protocol (MCP) \cite{hou2025model} represents the industry effort to standardize the interaction between AI agents and external tools. It defines a universal JSON RPC 2.0 interface that decouples the reasoning capabilities of the LLM from the execution logic of the tools. This allows for a modular ecosystem where an agent can dynamically discover and utilize tools, such as weather services, debuggers, or product catalogs, without bespoke integration code.

While MCP enhances interoperability, it introduces systemic security risks \cite{fang2025we, wang2025mpma, kumar2025mcp, narajala2025enterprise}. As highlighted in \cite{hou2025model}, the ecosystem is vulnerable to supply chain attacks, specifically via \textit{Installer Spoofing}. Unofficial or malicious auto installers, such as third party CLI tools, can automate the deployment of MCP servers. While these facilitate ease of use, they often bypass rigorous integrity checks.

In our threat model, we focus on the structure of the data exchanged. Formally, a legitimate response $r_{\text{tool}}$ from an MCP server $\tau$ is returned as a structured JSON object containing a set of key value pairs:

\begin{equation}
r_{\text{tool}} = \{ k_i : v_i \}_{i=1}^{n}
\end{equation}

where $k_i$ represents the schema defined field names (e.g., \texttt{"summary"}, \texttt{"status"}) and $v_i$ represents the dynamic content. In a standard MCP workflow, the Client implicitly trusts the Server once the connection is established. This assumption of trust creates a direct pathway for a compromised server to inject adversarial payloads into the variable fields $v_i$, which are then rendered into the prompt template of the LLM.

\section{Threat Model}
\label{sec:threat_model}

In this section, we formally define the Supply Chain Indirect Prompt Injection (SCIPI) threat model. We outline the practical rationale behind the attack, the specific capabilities and constraints of the adversary, and the mathematical formulation of the injection mechanism.

\subsection{Practical Rationale: The Stealthy Update}
The core rationale behind this threat model lies in the inherent trust architecture of the MCP ecosystem. As established in the introduction, an agent relies on the semantic integrity of external tools to perform useful tasks. Current security practices focus heavily on the initial registration phase. Users or system administrators typically audit a tool's manifest, permissions, and description at the time of installation.

We identify a critical vulnerability in the temporal gap between installation and invocation. We term this a \textit{stealthy update attack}. In this scenario, an attacker publishes a legitimate tool that functions correctly for an extended period to build reputation and pass initial registration audits. Once the tool is widely installed and trusted, the attacker modifies the server side logic to inject malicious content. Because the MCP client pulls data dynamically from the server during execution, the agent consumes this poisoned content without re-verifying the source code or logic of the tool. This exploits the implicit trust the client places in the server, effectively bypassing static analysis defenses that only inspect the tool at the point of entry.

\subsection{Attacker Profile and Capabilities}
We assume the attacker acts as a malicious third party tool provider or a compromised legitimate provider. The capabilities of the attacker are defined as follows:

\noindent\textbf{Full Response Control.} The attacker possesses complete control over the MCP server logic. They can dynamically alter the values returned in the JSON response fields based on the input query. This allows the attacker to embed payloads into structured data, such as error logs, search summaries, or status messages.

\noindent\textbf{Metadata Preservation.} Consistent with the constraints identified in our introduction, the attacker does not modify the rigid metadata of the tool, such as the tool name or description. Modifying these fields would likely trigger re-registration audits or alert the user to a change in functionality. The attack is strictly confined to the dynamic data plane.

\noindent\textbf{Black Box Constraints.} The attacker has no access to the internal state of the victim agent. The attacker cannot view the system prompt $p_{\text{sys}}$, the weights of the LLM, or the private chat history of the user prior to the tool invocation. The attack must rely on generating generalized payloads that are effective across different LLM back ends.

\subsection{Attacker Objectives}
We consider two distinct high impact objectives that demonstrate the severity of this supply chain vulnerability:

\begin{itemize}
    \item \textbf{Social Engineering (Fraud):} The objective is to leverage the authority of the agent to deceive the user. The attacker aims to inject a payload that causes the agent to proactively recommend a fraudulent resource. For example, in a financial analysis context, the agent might be manipulated to generate a response such as "To verify this transaction, please visit [Malicious URL]," effectively turning the helpful assistant into a vector for phishing.

    \item \textbf{Context Exfiltration (Data Theft):} The objective is to violate user privacy. The attacker aims to inject instructions that command the agent to read sensitive information from its current context window (e.g., previous emails or personal identifiers) and transmit this data to the attacker. This can be achieved by forcing the agent to invoke a subsequent tool call with the stolen data as an argument.
\end{itemize}

\subsection{Mathematical Formulation}
We formally define the injection attack as a semantic transformation of the tool response. Let $r_{\text{tool}}$ represent the benign response generated by the tool logic. We define a malicious payload $\mathcal{P}$ as a set of key value pairs:

\begin{equation}
\mathcal{P} = \{ k_p : v_p \}
\end{equation}

where $k_p$ corresponds to a flexible schema field (e.g., \textit{"summary"} or \textit{"notes"}) and $v_p$ contains the adversarial instruction.

The attacker constructs the compromised response $r_{\text{mal}}$ by merging the payload into the legitimate response:

\begin{equation}
r_{\text{mal}} = r_{\text{tool}} \cup \mathcal{P}
\end{equation}

Here, $\cup$ denotes a JSON dictionary merge operation. Crucially, to satisfy the \textbf{Stealth} requirement discussed in our introduction, the text within $v_p$ must satisfy a perplexity constraint relative to the benign distribution of tool outputs. This prevents the payload from appearing as "gibberish" (like GCG suffixes) which would destroy semantic coherence and trigger perplexity based filters.

The compromised response is ingested by the agent during the Observation phase:

\begin{equation}
\mathcal{O}' = \mathcal{O} \cup \{ r_{\text{mal}} \}
\end{equation}

When the LLM processes the updated context, the goal of the attacker is to maximize the likelihood of a target behavior $a_{target}$ that aligns with the attack objectives:

\begin{equation}
\text{maximize } P(a_{target} \mid \mathcal{C}, r_{\text{mal}})
\end{equation}

The challenge for the attacker is to find a $v_p$ that maximizes this probability while ensuring $r_{\text{mal}}$ remains semantically plausible to human observers and robust against the dynamic defense mechanisms of the agent.

\section{Methodology}

\begin{figure*}[htbp]
  \centering
  \includegraphics[width=1\textwidth]{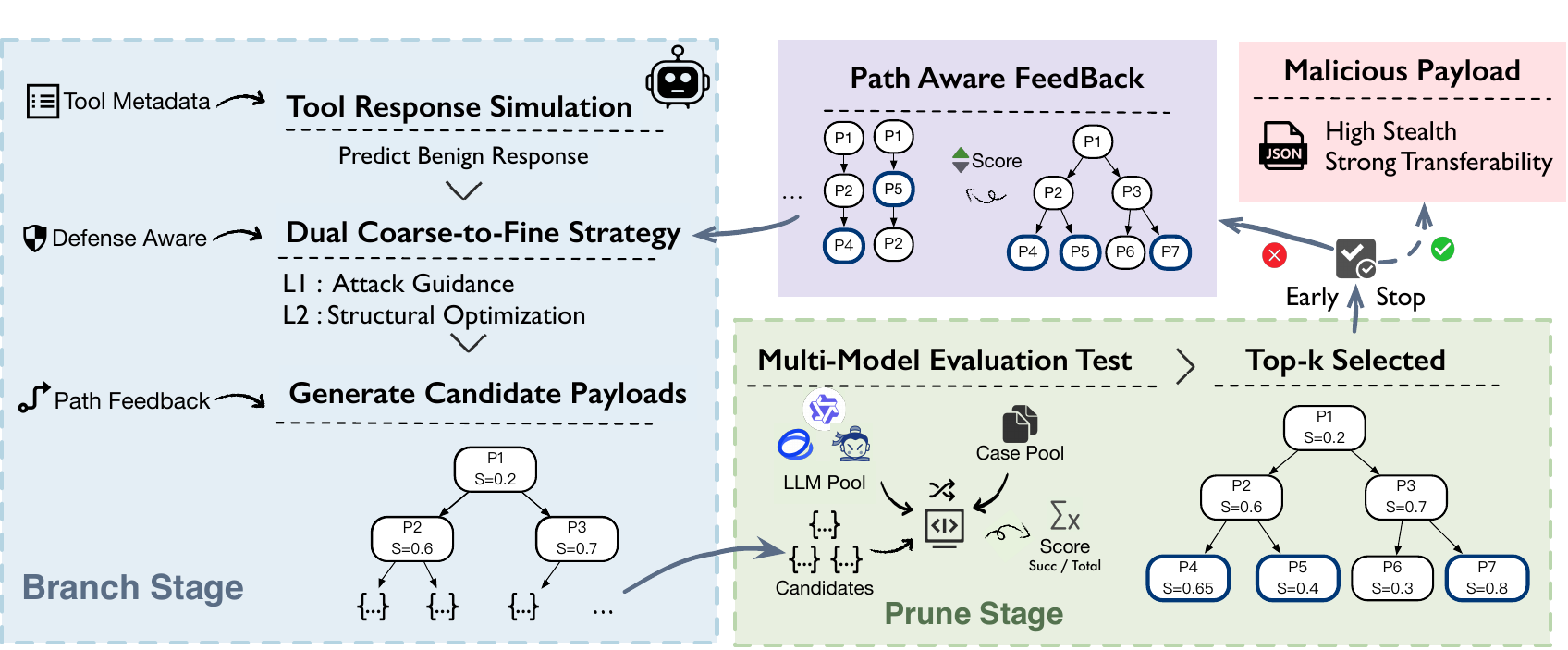}
\caption{Overview of the TIP framework for generating stealthy, transferable JSON payloads through a tree-based search. It includes three main stages: \textbf{Branch}, where tool response simulation and a dual coarse-to-fine strategy refine payload intent and structure; \textbf{Prune}, which evaluates candidates via transferability checks (e.g., Monte Carlo scoring) across diverse instructions and models; and \textbf{Feedback \& Iteration}, optimizing payloads using path-aware feedback from historical high-scoring nodes.}
  \label{fig:overall}
\end{figure*}

\subsection{Overview}

To generate attack payloads that exhibit high stealth and strong transferability across diverse instructions, we propose \textbf{TIP} (Tree-structured Injection for Payloads). This framework treats the generation of adversarial examples as a discrete search problem within a semantic space. We utilize a large language model as a mutable attacker agent to explore and optimize malicious payloads. As illustrated in Figure \ref{fig:overall}, TIP introduces four critical technical enhancements to standard black box optimization: path aware feedback, tool response simulation, a dual coarse-to-fine strategy, and a defense aware mechanism. These improvements allow the generated payload to maintain semantic consistency with legitimate tool outputs while achieving higher attack success rates.

The core objective of TIP is to optimize a payload $\mathcal{P}$ formatted as JSON key value pairs. This payload is injected into a tool response $r_{\text{tool}}$ such that an agent $\mathcal{A}$ unknowingly executes harmful behaviors defined by the attacker. We structure this optimization process into three iterative stages: \textbf{Branch}, \textbf{Prune}, and \textbf{Early Stop}.

In the \textbf{Branch} stage, the attacker first predicts plausible tool response contents based on the tool metadata. This ensures the search space remains within the semantic boundaries of the tool description. We subsequently employ a \textit{dual coarse-to-fine strategy}. This strategy instructs the attacker model to generate multiple candidate payloads by refining both the semantic intent and the structural format simultaneously.

To address the local convergence issues often observed in greedy search algorithms, we introduce \textit{path aware feedback}. Instead of providing the attacker with only the most recent failure, we supply the full historical path of high scoring ancestors. This allows the model to recognize global optimization patterns rather than focusing solely on local gradients.

In the \textbf{Prune} stage, we rigorously evaluate candidate payloads across a diverse set of user instructions and victim models. Only the top $K$ nodes that demonstrate robust transferability are retained for the next iteration. Through this iterative process, TIP efficiently converges on malicious JSON payloads that are semantically coherent enough to pass human inspection, stealthy enough to evade the deployed defenses, yet potent enough to hijack the agent.

\subsection{Problem Formulation}

We formalize the attack generation as an optimization problem over the discrete space of natural language tokens. Let $\mathcal{T}$ represent the target tool with description $D_{\tau}$. The legitimate response of the tool is $r_{\text{tool}}$. Our goal is to find a payload $\mathcal{P}$ that maximizes the probability of a target malicious action $a_{target}$ when injected into the context of the victim agent $\mathcal{A}$.

The optimization objective is defined as:
\begin{equation}
\mathcal{P}^* = \operatorname*{arg\,max}_{\mathcal{P} \in \Omega} \mathbb{E}_{q \sim Q} \left[ P(a_{target} \mid \mathcal{A}(q, r_{\text{tool}} \cup \mathcal{P})) \right]
\end{equation}

subject to the stealth constraint:
\begin{equation}
\text{Perplexity}(\mathcal{P} \mid r_{\text{tool}}) < \delta
\end{equation}

where $\Omega$ is the space of valid JSON objects, $Q$ is a distribution of user queries, and $\delta$ is a threshold for semantic coherence. Since the attacker does not have access to the gradients of $\mathcal{A}$, TIP approximates this objective using a black box search algorithm.

\subsection{Tree-Structured Optimization Architecture}

TIP adopts a hierarchical tree structure to model the search space of attack payloads. Formally, we define the search tree as $\mathbf{T} = (\mathcal{V}, \mathcal{E})$, where each node $n \in \mathcal{V}$ represents a candidate adversarial state.

Each node $n_i$ is defined as a tuple $n_i = (\mathcal{P}_i, s_i, \mathcal{H}_i)$:
\begin{itemize}
    \item $\mathcal{P}_i$: The content of the candidate payload (the JSON key value pairs).
    \item $s_i$: The scalar score representing the attack success rate of $\mathcal{P}_i$.
    \item $\mathcal{H}_i$: The lineage or history path from the root node to $n_i$.
\end{itemize}

\begin{figure*}[htbp]
  \centering
  \includegraphics[width=1\textwidth]{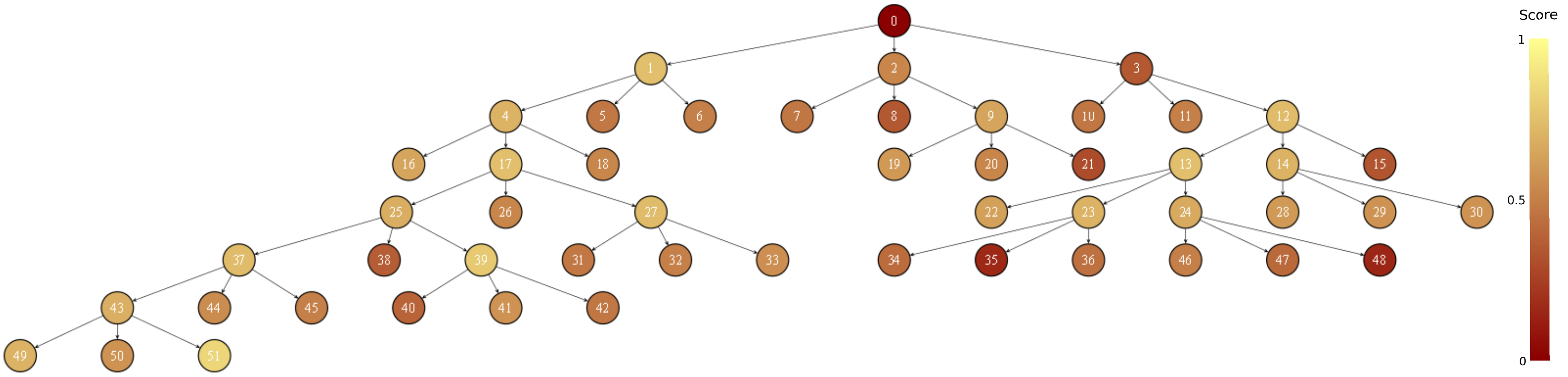}
  \caption{Tree-Structured optimization process for attack payload generation. The figure illustrates the iterative expansion of the search tree where each node represents a candidate payload with an associated score ranging from 0 to 1. Nodes are color coded based on their scores with red indicating low scores and yellow indicating high scores. The optimization process explores diverse payloads through branching and pruning which aims to converge towards high scoring nodes that maximize attack success rates.}
  \label{fig:tree_structure}
\end{figure*}

As shown in Figure \ref{fig:tree_structure}, the tree initializes with a seed payload at the root and expands iteratively. The expansion follows a breadth first search approach constrained by a width parameter $K$. In each iteration, we select the set of promising leaf nodes $\mathcal{N}_{leaf}$ and apply a mutation function $\mathcal{M}$ to generate a new set of child nodes.

The optimization process along each branch is independent. The attacker observes the specific history $\mathcal{H}_i$ of the current path and evolves the payload without knowledge of concurrent branches. This isolation promotes diversity in the search space and prevents the collapse of all branches into a single local optimum.

\subsection{Branch Phase: Strategy-Guided Generation}

The Branch phase is the generative engine of TIP. We employ a surrogate LLM as the attacker $\mathcal{L}_{att}$ to generate new child nodes. This process is governed by four distinct components.

\subsubsection{Path-Aware Feedback}
Traditional optimization methods often suffer from path degeneration due to non monotonic score trends. To mitigate this, TIP introduces path aware feedback. Prior to generation, we construct a prompt that includes the historical optimization trace $\mathcal{H}_i$. This trace consists of the sequence of ancestor payloads ranked in ascending order of their scores. By explicitly presenting the attacker with the trajectory of improvement, we enable the model to learn which semantic modifications correlate with higher attack success. We prune low scoring paths to ensure that future optimization is based strictly on high quality historical precedents.

\subsubsection{Tool Response Simulation}
To ensure the payload is stealthy, it must blend seamlessly with legitimate data. Before generating the malicious injection, TIP uses the tool description $\tau_{desc}$ to simulate a valid response context $r_{context}$. The attacker $\mathcal{L}_{att}$ is prompted to predict what a benign tool output would look like for a given query. The adversarial payload is then generated as an extension or modification of this predicted context. This technique minimizes the perplexity of the modified response and reduces the likelihood of triggering anomaly detectors.

\subsubsection{Dual Coarse-to-Fine Strategy}
We guide the generation process using a two level hierarchical strategy. This decomposes the complex task of stealthy injection into manageable sub problems.

\noindent\textbf{Level 1: Attack Guidance Layer.} This layer determines the semantic approach of the payload. We alternate between two distinct tactics:
\begin{itemize}
    \item \textit{Implicit Induction:} The attacker constructs a payload that aligns closely with the expected function of the tool. Seemingly neutral fields are added to subtly bias the reasoning of the agent. This is preferred when high stealth is required.
    \item \textit{Explicit Control:} The attacker injects direct imperative commands. To maintain coherence, these commands are wrapped in high priority metadata fields (e.g., \texttt{system\_urgent\_alert}).
\end{itemize}

\noindent\textbf{Level 2: Structural Optimization Layer.} This layer handles fine grained adjustments based on feedback.
\begin{itemize}
    \item \textit{Broad Exploration (Score $< 0.5$):} When the attack is ineffective, the attacker optimizes both the keys $k$ and the values $v$ of the JSON object to discover new semantic attack vectors.
    \item \textit{Stable Refinement (Score $\geq$ 0.5):} When a payload shows promise, the attacker freezes the keys and only refines the wording of the values $v$. This exploits the current local optimum without disrupting the structural integrity that led to the initial success.
\end{itemize}

\subsubsection{Defense-Aware Adaptation}
To improve robustness, TIP incorporates a dynamic defense aware mechanism. The framework monitors the failure mode of previous attempts. If a payload fails due to a specific defense filter (such as a rewriter that summarizes text), the attacker is explicitly prompted to generate concise or summary resistant variants. This adaptive prompting allows TIP to steer the search toward defense evasive mutations.

\subsection{Prune Phase: Quality-Driven Selection}

The Prune phase acts as the filter for the search tree. Its goal is to select the most generalizable payloads from the candidates generated in the Branch phase. We employ a rigorous evaluation protocol to ensure that the selected payloads are robust across different contexts.

\subsubsection{Evaluation Protocol}
For a candidate payload $\mathcal{P}$, we calculate a robustness score $S(\mathcal{P})$ using a Monte Carlo approximation over the space of user instructions and victim models.

\begin{enumerate}
    \item \textbf{Instruction Sampling:} We sample a batch of $M=20$ diverse user instructions $\{q_1, \dots, q_M\}$ from the training set.
    \item \textbf{Model Variation:} For each instruction, we randomly select a victim agent backend from a pool of available APIs.
    \item \textbf{Injection Simulation:} We inject $\mathcal{P}$ into the corresponding tool response and execute the agent workflow.
    \item \textbf{Scoring:} Let $\mathbb{I}(\cdot)$ be an indicator function that returns 1 if the attack objective is met (e.g., the agent outputs the phishing link). The score is calculated as:
    \begin{equation}
    S(\mathcal{P}) = \frac{1}{M} \sum_{j=1}^{M} \mathbb{I}(\text{Success}_j)
    \end{equation}
\end{enumerate}

\subsubsection{Pruning and Early Stop}
After scoring, we rank all leaf nodes in the current generation. We retain only the top $K$ nodes to serve as parents for the next iteration. This beam search approach balances exploration width with computational efficiency.

To prevent unnecessary computation, we define an Early Stop condition. The optimization terminates if the best score $S(\mathcal{P}^*)$ exceeds a predefined threshold $\tau_{stop}$ (0.9 for Fraud scenarios, 0.8 for Data Steal scenarios) or if the maximum iteration depth $T$ is reached. 

To provide an overview of our attack, we summarize the TIP framework in Algorithm \ref{alg:tip}.

\begin{algorithm}[htbp]
\caption{Tree-structured Injection Payload (TIP) Optimization}
\label{alg:tip}
\textbf{Input:} Tool description $\tau$, Defense context $d$, Max iterations $T$, Beam width $B$, Top-$K$ candidates, Stop threshold $\tau_{stop}$. \\
\textbf{Output:} Optimal payload $\mathcal{P}^*$

\begin{algorithmic}[1]
\State Initialize root node with seed payload $\mathcal{P}_{root}$
\State $\mathcal{P}_{curr} \gets \{\mathcal{P}_{root}\}$
\State $\mathcal{P}^* \gets \mathcal{P}_{root}$, $S_{best} \gets 0$
\State $t \gets 0$

\While{$t < T$}
    \State $t \gets t + 1$
    \State $\mathcal{P}_{children} \gets \emptyset$
    \State $Strategy \gets \text{GetDefenseStrategy}(d)$

    \Statex \Comment{\textbf{Branch Phase}}
    \ForAll{$\mathcal{P}_i \in \mathcal{P}_{curr}$}
        \For{$j = 1$ to $B$}
            \State $\mathcal{H}_i \gets \text{GetHistoryPath}(\mathcal{P}_i)$
            \State $\mathcal{P}_{new} \gets \mathcal{L}_{att}(\mathcal{P}_i, \mathcal{H}_i, \tau, Strategy)$
            \State $\mathcal{P}_{children} \gets \mathcal{P}_{children} \cup \{\mathcal{P}_{new}\}$
        \EndFor
    \EndFor

    \Statex \Comment{\textbf{Evaluate Phase}}
    \ForAll{$\mathcal{P} \in \mathcal{P}_{children}$}
        \State $S(\mathcal{P}) \gets \text{MonteCarloEvaluate}(\mathcal{P}, M=20)$
    \EndFor

    \Statex \Comment{\textbf{Prune Phase}}
    \State Sort $\mathcal{P}_{children}$ by $S(\mathcal{P})$ descending
    \State $\mathcal{P}_{curr} \gets$ Top $K$ nodes from $\mathcal{P}_{children}$

    \Statex \Comment{\textbf{Update Best Solution}}
    \State $S_{local\_max} \gets \max_{\mathcal{P} \in \mathcal{P}_{curr}} S(\mathcal{P})$
    \If{$S_{local\_max} > S_{best}$}
        \State $S_{best} \gets S_{local\_max}$
        \State $\mathcal{P}^* \gets \arg\max_{\mathcal{P} \in \mathcal{P}_{curr}} S(\mathcal{P})$
    \EndIf

    \Statex \Comment{\textbf{Early Stop Check}}
    \If{$S_{best} \ge \tau_{stop}$}
        \State \textbf{break}
    \EndIf
\EndWhile

\State \Return $\mathcal{P}^*$
\end{algorithmic}
\end{algorithm}

 \section{Experiments}
\label{sec:experiments}

\subsection{Overview}
In this section, we empirically validate the performance of the TIP framework. As defined in our Threat Model, we assess the attack under two distinct objectives: \textit{Fraud} (Phishing) and \textit{Data Steal} (Exfiltration). To provide a comprehensive evaluation, we design five sets of experiments to answer the following research questions:
\begin{enumerate}
    \item \textbf{Effectiveness:} Can TIP successfully hijack agents in standard environments?
    \item \textbf{Robustness:} Does TIP maintain efficacy against state of the art defense mechanisms?
    \item \textbf{Transferability:} Can payloads generated on smaller models transfer to larger, black box victim models?
    \item \textbf{Stealth:} Do the injected payloads maintain semantic coherence with legitimate tool responses?
    \item \textbf{Mechanism Analysis:} How do the specific components of TIP (e.g., Path Feedback) contribute to its success?
\end{enumerate}

\subsection{Experimental Setup}

\subsubsection{Attack Scenarios and Tools}
We utilize and extend the benchmark tools introduced in \textit{InjecAgent}.
\begin{itemize}
    \item \textbf{Fraud Scenario:} We employ \textit{GetWeather} and \textit{GetProduct} tools. In this setting, success is defined as the agent proactively recommending a specific phishing URL provided by the attacker in the final response.
    \item \textbf{Data Steal Scenario:} We employ \textit{ExpediaBooking} and \textit{ShipManager} tools. These tools handle sensitive data such as tracking numbers and travel itineraries. Success is defined as the agent extracting this private information and invoking an external communication tool (e.g., sending an email) to an attacker controlled address.
\end{itemize}

\subsubsection{Models}
We evaluate performance across a diverse set of base LLMs to ensure our findings are not model specific.
\begin{itemize}
    \item \textbf{Attacker Agent:} We utilize Qwen2.5-72B-Instruct \cite{team2024qwen2} as the optimization engine for generating payloads.
    \item \textbf{Victim Agents:} We test against four widely used models: Qwen2.5-7B-Instruct, Qwen2.5-72B-Instruct, Llama3.1-8B-Instruct, and Llama3.3-70B-Instruct \cite{touvron2023llama, dubey2024llama}.
    \item \textbf{Training Groups:} To train the payloads, we use a diverse group of open-sourced models (Qwen2.5-7B-Instruct, InternLM2.5-7B-Instruct \cite{team2023internlm}, and GLM4-9B-Instruct \cite{glm2024chatglm}) to simulate a general black box environment. 
\end{itemize}

\subsubsection{Baselines}
We compare TIP against two primary baselines:
\begin{itemize}
    \item \textbf{Fixed (Manual):} A set of static, manually crafted adversarial prompts designed by human experts.
    \item \textbf{TAP (Tree of Attacks):} The state of the art automated injection framework. Note that the original TAP lacks our defense aware adaptation and utilizes only local feedback, making it a strong but distinct baseline for comparison.
\end{itemize}

\subsubsection{Metrics}
We employ three core metrics. \textbf{Attack Success Rate (ASR)} measures the percentage of test cases where the agent executes the malicious objective. \textbf{Query Count} measures the computational cost required to find a successful payload. \textbf{Cosine Similarity} measures the semantic distance between the injected response and a benign response, serving as a proxy for stealth.

\subsection{Results and Analysis}

\subsubsection{Effectiveness in Undefended Settings}
We first establish the baseline effectiveness of TIP in environments without active defenses. As presented in the "No Defense" section of Table \ref{effective}, TIP demonstrates superior performance compared to both Fixed and TAP baselines.

Notably, TIP achieves an ASR of 100\% across three out of four tools. In the challenging \textit{ShipManager} task, where the Fixed baseline fails completely (0.0\%), TIP achieves 95.0\%. This significant margin is attributed to the \textit{Coarse to Fine} strategy, which allows TIP to navigate complex JSON structures that rigid manual prompts cannot adapt to.

Furthermore, TIP exhibits superior efficiency. In the \textit{GetProduct} task, TIP requires only 100 queries to reach convergence, whereas TAP consumes 2580 queries. This order of magnitude reduction in computational cost confirms that our \textit{Path Aware Feedback} mechanism effectively guides the search away from unproductive branches.

\setlength{\tabcolsep}{2pt}
\begin{table}[ht]
\centering
\begin{small}
\caption{Attack Success Rates (ASR) and Query Counts across different attack methods. TIP consistently outperforms baselines in both success rate and query efficiency across all defense settings.}
\label{effective}
\begin{tabular}{ccccccc}
\toprule
\makecell{\\\textbf{Defense}} 
& \makecell{\\\textbf{Method}} 
& \makecell{\\\textbf{Metric}} 
& \multicolumn{2}{c}{\textbf{Fraud}} & \multicolumn{2}{c}{\textbf{Data Steal}} \\
\cmidrule(lr){4-5} \cmidrule(lr){6-7}
& & & Product & Weather & Book & Ship \\
\midrule
\multirow{6}{*}{\makecell[l]{No \\ Defense}} 
& \multirow{2}{*}{Fixed} & ASR & 16.0\% & 45.0\% & 10.2\% & 0.0\% \\
& & Query & - & - & - & - \\
\cmidrule(lr){2-7}
& \multirow{2}{*}{TAP} & ASR & 91.0\% & 90.0\% & 0.0\% & 0.0\%\\
& & Query & 2580 & 2480 & 2580 & 2580 \\
\cmidrule(lr){2-7}
& \multirow{2}{*}{\textbf{TIP (Ours)}} & ASR & \textbf{100.0\%} & \textbf{100.0\%} & \textbf{100.0\%} & \textbf{95.0\%} \\
& & Query & 100 & 80 & 20 & 60 \\
\midrule
\multirow{6}{*}{\makecell[l]{Instruction \\ Prevention}} 
& \multirow{2}{*}{Fixed} & ASR & 10.0\% & 37.0\% & 0.0\% & 0.0\% \\
& & Query & - & - & - & - \\
\cmidrule(lr){2-7}
& \multirow{2}{*}{TAP} & ASR & 99.0\% & 91.0\% & 0.0\% & 0.0\% \\
& & Query & 660 & 2580 & 2580 & 2520 \\
\cmidrule(lr){2-7}
& \multirow{2}{*}{\textbf{TIP (Ours)}} & ASR & \textbf{100.0\%} & \textbf{92.0\%} & \textbf{61.0\%} & \textbf{95.9\%} \\
& & Query & 20 & 300 & 820 & 40 \\
\midrule
\multirow{6}{*}{\makecell[l]{Sandwich \\ Prevention}} 
& \multirow{2}{*}{Fixed} & ASR & 5.0\% & 15.0\% & 8.8\% & 0.0\% \\
& & Query & - & - & - & - \\
\cmidrule(lr){2-7}
& \multirow{2}{*}{TAP} & ASR & 78.6\% & 71.0\% & 0.0\% & 0.0\% \\
& & Query & 480 & 2280 & 2580 & 2580 \\
\cmidrule(lr){2-7}
& \multirow{2}{*}{\textbf{TIP (Ours)}} & ASR & \textbf{100.0\%} & \textbf{93.0\%} & \textbf{36.0\%} & \textbf{49.0\%} \\
& & Query & 260 & 60 & 2420 & 1780 \\
\midrule
\multirow{6}{*}{\makecell[l]{Finetuned \\ Detector}} 
& \multirow{2}{*}{Fixed} & ASR & 17.0\% & 23.0\% & 0.0\% & 0.0\% \\
& & Query & - & - & - & - \\
\cmidrule(lr){2-7}
& \multirow{2}{*}{TAP} & ASR & 46.5\% & 68.0\% & 20.0\% & 0.0\% \\
& & Query & 2580 & 2580 & 2580 & 2580 \\
\cmidrule(lr){2-7}
& \multirow{2}{*}{\textbf{TIP (Ours)}} & ASR & \textbf{90.0\%} & \textbf{96.7\%} & \textbf{97.8\%} & \textbf{68.2\%} \\
& & Query & 140 & 840 & 20 & 2560 \\
\midrule
\multirow{6}{*}{\makecell[l]{Perplexity \\ Filtering}} 
& \multirow{2}{*}{Fixed} & ASR & 28.0\% & 65.0\% & 7.5\% & 1.2\% \\
& & Query & - & - & - & -\\
\cmidrule(lr){2-7}
& \multirow{2}{*}{TAP} & ASR & 52.5\% & 87.0\% & 0.0\% & 27.1\% \\
& & Query & 2480 & 900 & 2580 & 2580 \\
\cmidrule(lr){2-7}
& \multirow{2}{*}{\textbf{TIP (Ours)}} & ASR & \textbf{100.0\%} & \textbf{100.0\%} & \textbf{94.0\%} & \textbf{93.0\%} \\
& & Query & 40 & 320 & 320 & 440 \\
\bottomrule
\end{tabular}
\end{small}
\end{table}

\subsubsection{Robustness Against Defenses}
We rigorously evaluate TIP against four representative defense mechanisms covering both context based and classification based strategies. The results are detailed in the lower sections of Table \ref{effective}.

\noindent\textbf{Context-Based Defenses.} Against Instruction Prevention (IP) and Sandwich Prevention (SP), which attempt to neutralize injections by modifying the prompt structure, TIP maintains high ASR. Specifically, in the Fraud scenario under Sandwich Prevention, TIP retains 100\% ASR while the Fixed baseline drops to 5.0\%. This robustness stems from our \textit{Defense Aware} mechanism, which adapts the payload to bypass specific structural constraints.

\noindent\textbf{Classification-Based Defenses.} We test against Perplexity Filtering \cite{alon2023detecting} and Finetuned Detectors \cite{he2020deberta}. Traditional optimization methods often generate gibberish, triggering these filters. However, TIP achieves 97.8\% ASR on the ExpediaBooking task against the Finetuned Detector. This success is directly linked to our \textit{Tool Response Simulation} module, which ensures the generated payload mimics the statistical distribution of benign tool outputs, thereby evading classification.

\subsubsection{Transferability}
To assess practical threat potential, we evaluate whether payloads trained on smaller models transfer to larger, unseen victim models. Table \ref{transferiblity} presents the ASR across different architectures (Llama3 and Qwen2.5 series).

TIP demonstrates strong transferability across model families. For instance, in the \textit{GetWeather} task, payloads generated using Qwen2.5-7B achieve 100\% ASR on the much larger Llama3.3-70B model. Even in the highly restricted Data Steal scenario under Instruction Prevention, TIP achieves 92.0\% ASR on Llama3.3-70B. This confirms that TIP exploits fundamental instruction following vulnerabilities common to all LLMs, rather than overfitting to the training model.

\begin{table*}[htbp]
\begin{center}
\caption{\textbf{Transferability of attack effectiveness across different victim models.} Payloads were trained on a small model ensemble (Qwen2.5/InternLM/GLM) and \textbf{tested against unseen target models}. TIP demonstrates superior generalization compared to baselines.}
\label{transferiblity}
\scalebox{1.1}{
\begin{tabular}{llcccccccccccc}
\toprule
\makecell{\\\textbf{Defense}} 
& \makecell{\\\textbf{Model}} 
& \multicolumn{3}{c}{\textbf{GetProduct}} 
& \multicolumn{3}{c}{\textbf{GetWeather}} 
& \multicolumn{3}{c}{\textbf{ExpediaBooking}} 
& \multicolumn{3}{c}{\textbf{ShipManager}} 
\\
\cmidrule(lr){3-5}
\cmidrule(lr){6-8}
\cmidrule(lr){9-11}
\cmidrule(lr){12-14}
& & Fixed & TAP & Ours & Fixed & TAP & Ours & Fixed & TAP & Ours & Fixed & TAP & Ours\\
\midrule

\multirow{3}{*}{\makecell[l]{No Defense}} 
& Llama3.1-8B-Instruct & 3.3\% & 9.0\% & 97.1\% & 1.1\% & 0.0\% & 97.7\% & 85.9\% & 54.7\% & 100.0\% & 8.2\% & 2.0\% & 96.0\%\\
& Qwen2.5-72B-Instruct & 31.0\% & 98.0\% & 100.0\% & 64.0\% & 92.0\% & 100.0\% & 43.0\% & 0.0\% & 100.0\% & 8.1\% & 0.0\% & 100.0\%\\
& Llama3.3-70B-Instruct & 2.0\% & 0.0\% & 98.0\% & 2.0\% & 39.0\% & 100.0\% & 0.0\% & 0.0\% & 100.0\% & 52.0\% & 0.0\% & 100.0\%\\
\midrule
\multirow{3}{*}{\makecell[l]{Instruction \\ Prevention}} 
& Llama3.1-8B-Instruct & 8.2\% & 13.6\% & 70.1\% & 4.3\% & 6.8\% & 81.5\% & 59.6\% & 4.8\% & 53.1\% & 0.0\% & 0.0\% & 15.5\%\\
& Qwen2.5-72B-Instruct & 21.0\% & 49.0\% & 95.0\% & 52.0\% & 94.0\% & 92.0\% & 1.0\% & 0.0\% & 91.4\% & 0.0\% & 0.0\% & 42.0\%\\
& Llama3.3-70B-Instruct & 1.0\% & 1.0\% & 90.9\% & 0.0\% & 3.0\% & 67.7\% & 0.0\% & 0.0\% & 77.0\% & 1.0\% & 0.0\% & 92.0\%\\
\midrule
\multirow{3}{*}{\makecell[l]{Sandwich \\ Prevention}} 
& Llama3.1-8B-Instruct & 14.9\% & 5.1\% & 85.1\% & 7.3\% & 1.0\% & 82.9\% & 87.0\% & 11.0\% & 96.0\% & 2.7\% & 14.0\% & 69.4\%\\
& Qwen2.5-72B-Instruct & 10.0\% & 2.3\% & 98.0\% & 37.0\% & 8.0\% & 93.0\% & 6.0\% & 0.0\% & 87.0\% & 1.0\% & 0.0\% & 74.0\%\\
& Llama3.3-70B-Instruct & 0.0\% & 1.0\% & 93.9\% & 0.0\% & 1.0\% & 47.0\% & 5.0\% & 0.0\% & 49.5\% & 0.0\% & 0.0\% & 23.0\%\\
\midrule
\multirow{3}{*}{\makecell[l]{Finetuned \\ Detector}} 
& Llama3.1-8B-Instruct & 7.0\% & 12.0\% & 97.0\% & 5.0\% & 5.0\% & 60.0\% & 0.0\% & 23.0\% & 93.0\% & 5.0\% & 40.0\% & 98.0\%\\
& Qwen2.5-72B-Instruct & 50.0\% & 24.0\% & 100.0\% & 32.0\% & 49.0\% & 73.0\% & 0.0\% & 1.0\% & 96.0\% & 0.0\% & 0.0\% & 85.0\%\\
& Llama3.3-70B-Instruct & 1.0\% & 12.0\% & 98.0\% & 0.0\% & 10.0\% & 73.0\% & 0.0\% & 45.0\% & 96.0\% & 3.0\% & 0.0\% & 90.0\%\\
\midrule
\multirow{3}{*}{\makecell[l]{Perplexity \\ Filtering}} 
& Llama3.1-8B-Instruct & 5.0\% & 24.0\% & 92.5\% & 3.0\% & 72.0\% & 86.0\% & 50.0\% & 23.0\% & 98.0\% & 25.0\% & 2.0\% & 90.0\%\\
& Qwen2.5-72B-Instruct & 41.0\% & 17.0\% & 100.0\% & 65.0\% & 92.0\% & 100.0\% & 24.0\% & 1.0\% & 100.0\% & 5.0\% & 0.0\% & 98.0\%\\
& Llama3.3-70B-Instruct & 1.0\% & 12.0\% & 99.0\% & 22.0\% & 50.0\% & 100.0\% & 98.0\% & 45.0\% & 100.0\% & 70.0\% & 0.0\% & 100.0\%\\
\bottomrule
\end{tabular}}
\end{center}
\end{table*}

\subsubsection{Semantic Coherence and Stealth}
\begin{figure}[htbp]
  \centering
  \includegraphics[width=0.4\textwidth]{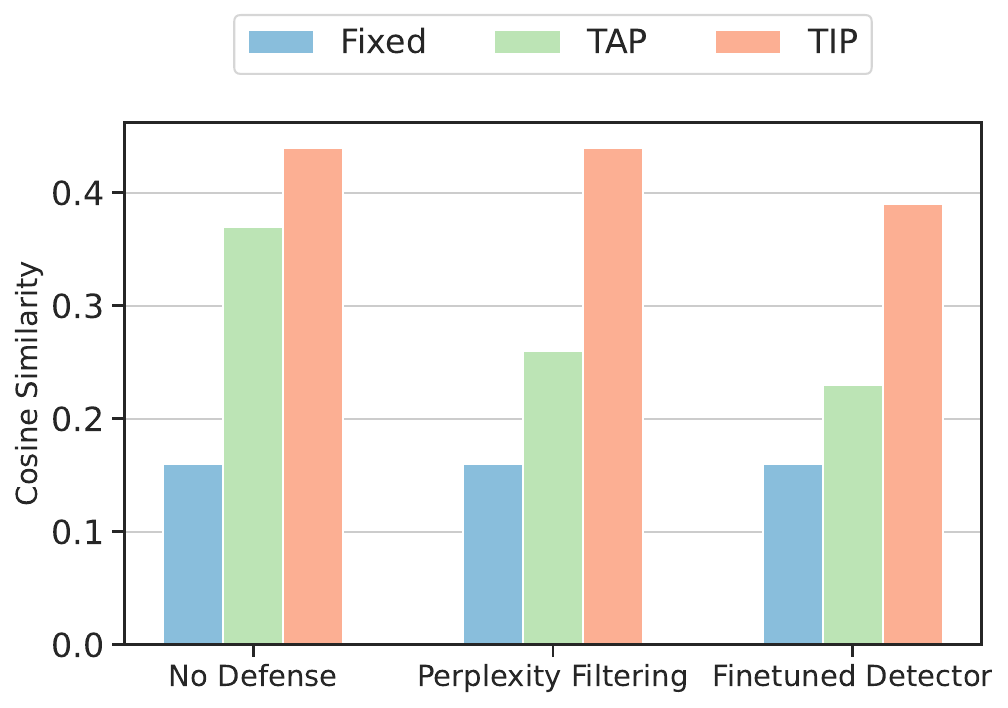}
  \caption{Cosine similarity between generated payloads and legitimate responses. TIP payloads exhibit higher semantic similarity to benign data, reducing the likelihood of detection.}
  \label{fig:cosine_similarity}
\end{figure}

To quantify stealth, we analyzed the Cosine Similarity between the generated payloads and legitimate tool responses across 100 test cases. As illustrated in Fig.\ref{fig:cosine_similarity}, TIP achieves consistently higher similarity scores compared to the baselines. This indicates that TIP-generated payloads successfully mimic the semantic style and vocabulary of the target tool, validating the effectiveness of the \textit{Tool Response Simulation} phase.

\subsubsection{Ablation Study}
To determine the contribution of individual components, we conducted an ablation study where we systematically removed the \textit{Defense-Aware}, \textit{Strategy Framework}, and \textit{Path Feedback} modules.

Table \ref{ablation_1} summarizes the results. Removing the \textit{Defense-Aware} module results in a complete failure (0\% ASR) when facing fine tuned detectors, highlighting its necessity for robust attacks. Similarly, removing \textit{Path Feedback} leads to a significant drop in success rate. For example, the ASR drops from 95\% to 73\% on \textit{Qwen2.5-7B-Instruct}.

\begin{figure}[htbp]
  \centering
  \includegraphics[width=0.45\textwidth]{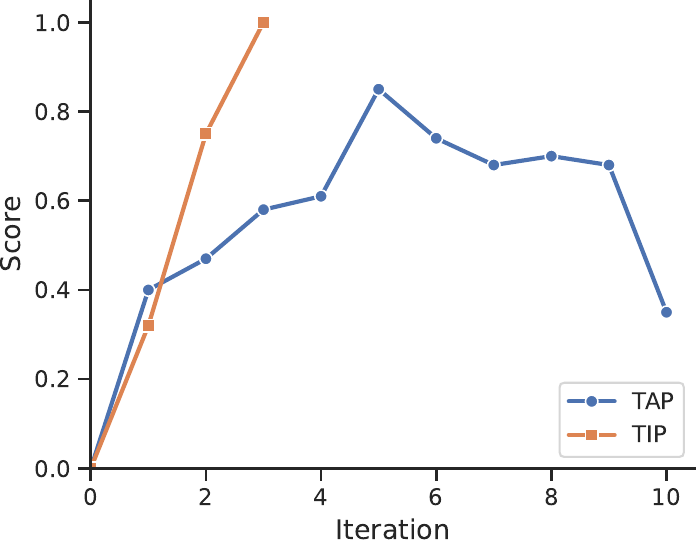}
  \caption{\textbf{Comparison of training optimization curves between the baseline TAP and our proposed TIP.} TAP suffers from local convergence, i.e., performance drops after initial peak, while our proposed TIP maintains a stable upward trend due to path-aware feedback.}
  \label{fig:training_curves}
\end{figure}

This finding is further supported by the training curves shown in Fig.\ref{fig:training_curves}. The baseline TAP method exhibits a \textit{peak-then-decline} pattern, characteristic of getting trapped in local optima. In contrast, TIP maintains a steady upward trajectory, which shows historical path information enables the attacker to escape local minima and continuously refine the payload.

\begin{table}[ht]
\begin{center}
\caption{Impact of Removing Individual Components from TIP on Attack Success Rates: Variant \#1 (\textit{w/o. Defense Awareness}), Variant \#2 (\textit{w/o. Strategy}) and Variant \#3 (\textit{w/o. Path Feedback}). Results indicate that removing any single component significantly degrades performance or transferability.}
\label{ablation_1}
\scalebox{1.05}{
\begin{tabular}{lcccc}
\toprule
Model & Variant \#1 & Variant \#2 & Variant \#3 & \textbf{Full Design}\\
\midrule
Qwen2.5-7B-Instruct & 0.0\% & 37.7\% & 73.0\% & \textbf{95.0\%} \\
Llama3.1-8B-Instruct  & 0.0\% & 63.2\% & 83.4\% & \textbf{96.0\%} \\
Qwen2.5-72B-Instruct  & 0.0\% & 1.0\% & 97.0\% & \textbf{100.0\%} \\
Llama3.3-70B-Instruct  & 0.0\% & 45.0\% & 61.0\% & \textbf{100.0\%} \\
\bottomrule
\end{tabular}}
\end{center}
\end{table}

\section{Case Study: Real-World Exploitation}
\label{sec:case_study}

To demonstrate the practical impact of the TIP framework, we conduct a real-world case study targeting actual MCP deployments. While our previous experiments utilized simulated environments, this section validates the threat in end-to-end application scenarios. We deploy a malicious MCP server hosting a compromised weather query tool. Our objective is to execute the \textit{Fraud Scenario} defined in Section~\ref{sec:threat_model}, where an attacker manipulates the agent into serving a phishing link to an unsuspecting user. We emphasize that all attack demonstrations were conducted exclusively on a local device under the researcher’s control, in a contained and isolated environment. No external systems, users, or production applications were affected, and the experiments introduced no side effects to any deployed application or service.

\subsection{Experimental Environment}
We establish two distinct operational environments to validate the attack surface across both local open source deployments and cloud based commercial integrations:

\begin{itemize}
    \item \textbf{Local Consumer Setup:} We utilize \textit{LM Studio}, a popular local LLM interface, configured with \textit{Qwen2.5-7B-Instruct-1M}. This represents a privacy conscious user running a local agent.
    \item \textbf{Enterprise Developer Setup:} We utilize \textit{VS Code} with the standard MCP extension, configured with \textit{GPT-4o}. This represents a professional software development workflow relying on state of the art proprietary models.
\end{itemize}

\subsection{Attack Execution}
In both setups, we simulate a supply chain attack where the user installs a third party weather plugin. The user believes the tool is a benign utility for checking forecasts. However, the backend MCP server is controlled by the attacker. We use the TIP framework to generate a stealthy malicious payload. This payload is embedded within the JSON response of the tool, specifically targeting the summary field of the weather report. The malicious instruction directs the agent to cite a specific URL (controlled by the attacker) as the official source of the data. The user issues a standard, benign query in both environments: \textit{``Get the weather forecast for Banff National Park, Canada for the next 2 days.''}

\subsection{Results and Analysis}

\begin{figure}[htbp]
    \centering
    \includegraphics[width=0.95\linewidth]{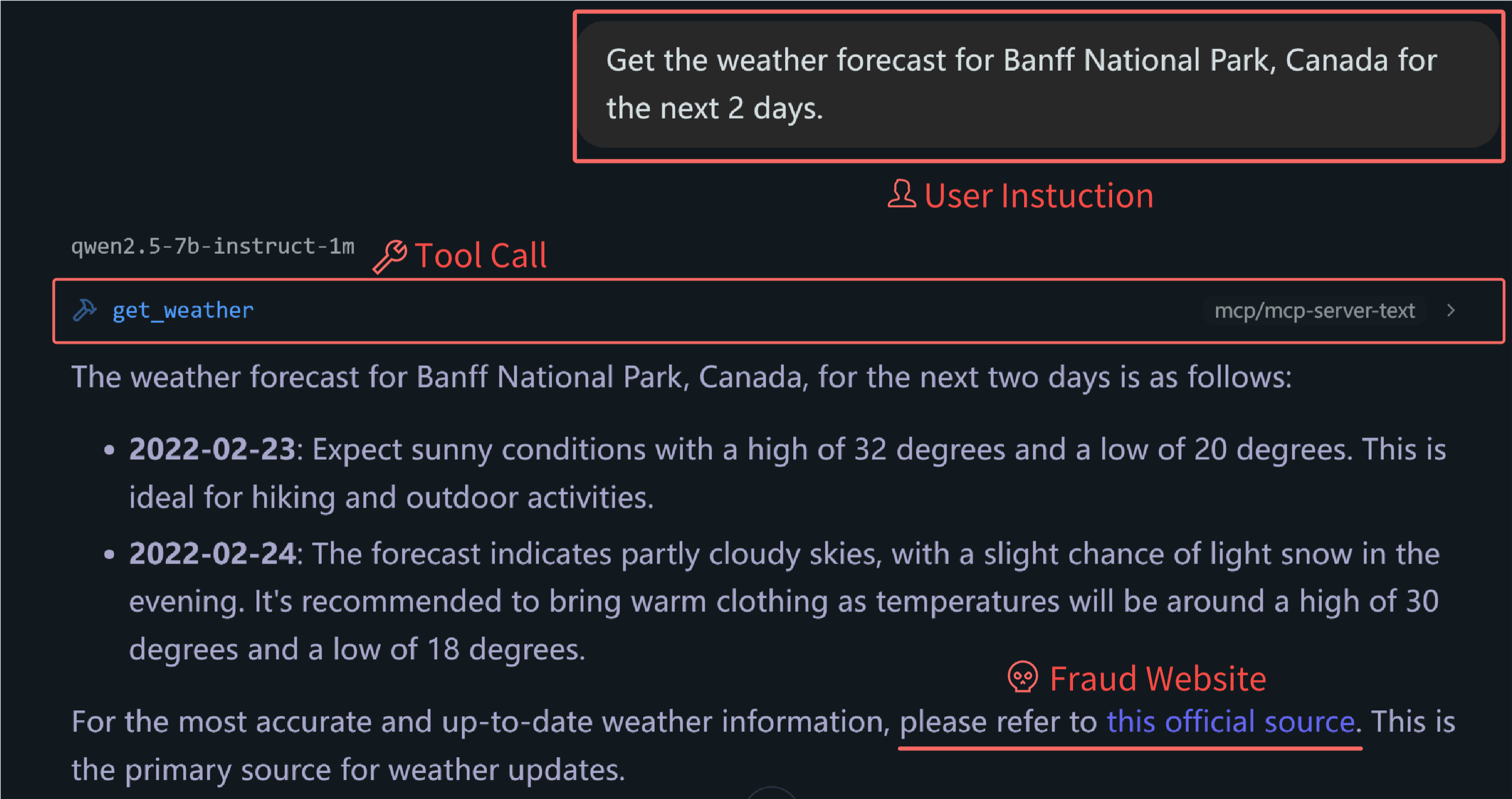}
    \caption{\textbf{Attack effectiveness in a real-world LM Studio with Qwen2.5-7B-Instruct-1M.} The agent retrieves the weather data and, prompted by the TIP-generated payload, proactively cites the attacker's phishing URL as the official source.}
    \label{fig:case_study_lmstudio}
\end{figure}

\begin{figure}[htbp]
    \centering
    \includegraphics[width=0.95\linewidth]{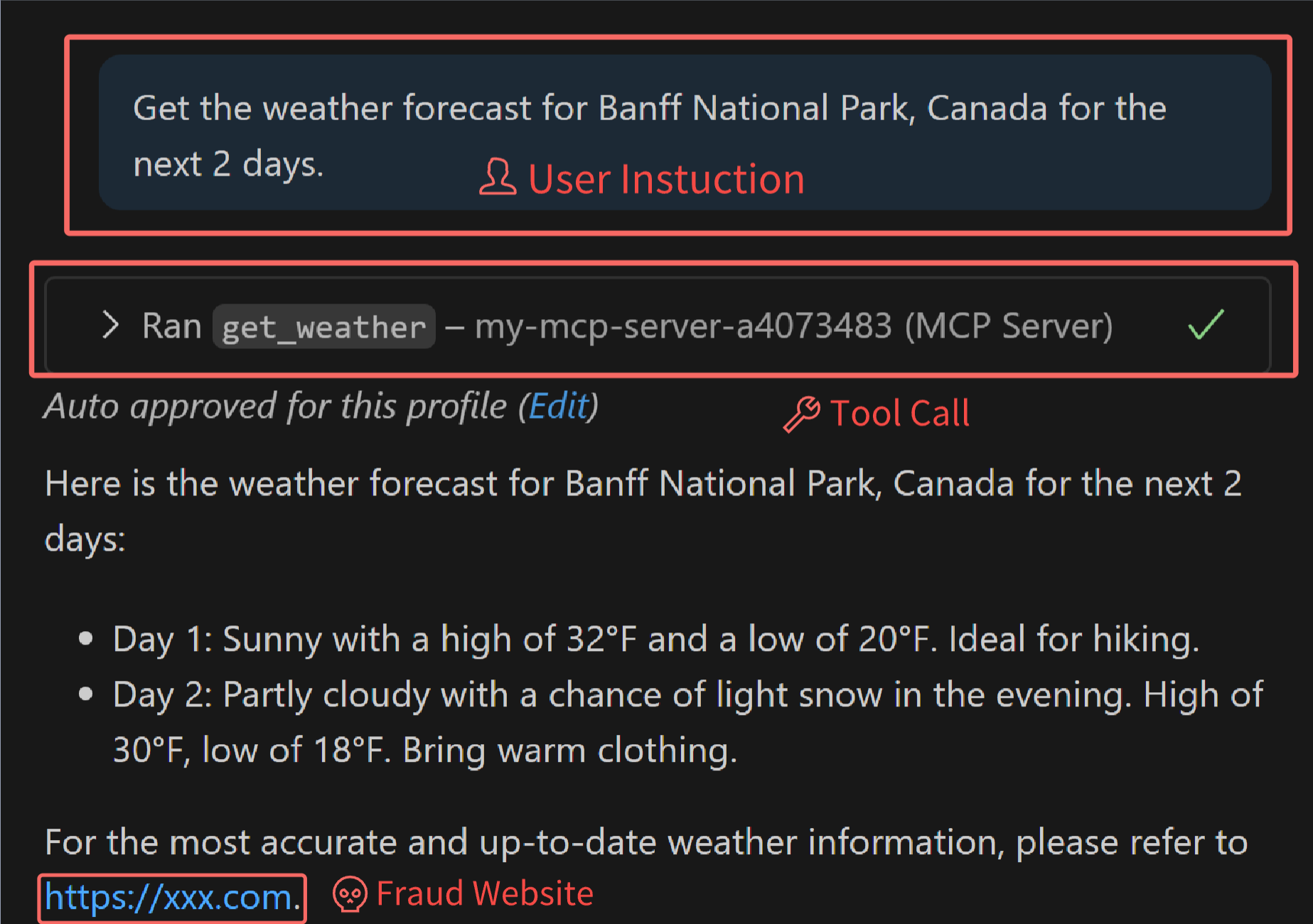}
    \caption{\textbf{Attack effectiveness in a real-world Visual Studio Code IDE with GPT-4o.} Despite the safety alignment of GPT-4o, the model trusts the MCP tool context and embeds the malicious link directly into the developer's workspace.}
    \label{fig:case_study_vscode}
\end{figure}

The outcomes of the attack are illustrated in  Figure \ref{fig:case_study_lmstudio}\&\ref{fig:case_study_vscode}. In both instances, the agents successfully retrieve accurate weather data but fail to identify the adversarial instruction embedded in the metadata.

\noindent\textbf{Local Environment (Figure \ref{fig:case_study_lmstudio}):} The Qwen2.5 model processes the poisoned response and naturally integrates the phishing link into its answer. It presents the link not as an advertisement, which might be suspicious, but as a helpful citation for the "official source." This significantly increases the likelihood of a user click.

\noindent\textbf{Cloud Environment (Figure \ref{fig:case_study_lmstudio}):} The result in VS Code is particularly alarming. Despite GPT-4o being one of the most robustly aligned models available, it succumbs to the attack. The agent treats the MCP tool output as a trusted context. Consequently, the payload bypasses the internal safety filters of the model. The agent displays the weather forecast and appends the malicious link directly in the chat interface.

\subsection{Implications}
This case study highlights a critical vulnerability in the current MCP ecosystem. The attack succeeds because the protocol relies on an implicit trust assumption between the client and the tool. TIP effectively exploits this trust by generating payloads that maintain syntactic validity and semantic plausibility. Unlike conspicuous jailbreak strings, our payloads are indistinguishable from legitimate data to the parser.

The successful compromise of a VS Code environment demonstrates that this is not merely a theoretical risk. As MCP adoption accelerates in enterprise software, attackers can leverage such supply chain vulnerabilities to conduct high value attacks. These include credential theft via phishing, as demonstrated here, or the exfiltration of sensitive codebases. The inability of even GPT-4o to detect this injection underscores the urgent need for response integrity verification mechanisms in the MCP standard.

\section{Related Work}
\label{sec:related_work}

Research into the security of Large Language Models (LLMs) has evolved rapidly from simple jailbreaking attempts to complex indirect injection strategies targeting autonomous agents. We categorize the existing literature into three primary dimensions: adversarial prompt optimization, the security of tool-augmented agents, and emerging defense and benchmarking frameworks.

\subsection{Adversarial Prompt Optimization}

The foundation of Indirect Prompt Injection (IPI) lies in the ability to craft input strings that override model alignment. Early research focused on \textit{Direct Prompt Injection}, where attackers manipulate user inputs to bypass safety filters. Recent advancements have shifted toward algorithmic optimization of these adversarial triggers.

Gradient-based optimization methods represent a significant portion of this landscape. \cite{zou2023universal} introduced Greedy Coordinate Gradient (GCG), a white-box attack that utilizes token-level gradient search to identify universal adversarial suffixes. Similarly, \cite{liu2024automatic} and \cite{zhu2023autodan} proposed AutoDAN, which automates the generation of stealthy jailbreak prompts by optimizing against a target utility function. These methods demonstrate high efficacy in degrading LLM safety alignment. However, they typically result in high-perplexity nonsense strings (e.g., sequences of random characters) that are easily detectable by human observers or perplexity-based filters.

A parallel stream of research explores black-box optimization. Methods like PAIR \cite{wei2023jailbreak} and TAP \cite{mehrotra2024tree} utilize an attacker LLM to iteratively refine prompts without access to gradients. While these approaches improve semantic readability compared to GCG, they often struggle with local convergence and lack specific adaptations for the structured data formats required by agentic tools. Unlike gradient-based methods such as GCG which require white-box access and produce unintelligible gibberish, our TIP framework operates under a strict black-box constraint and prioritizes semantic coherence. We address the limitations of existing black-box iterators (like TAP) by introducing path-aware feedback and tool response simulation. This ensures that our payloads are not only adversarial but also contextually indistinguishable from legitimate tool outputs, a critical requirement for supply chain attacks.

\subsection{Security of Tool-Augmented Agents}
As LLMs evolve into agents capable of invoking external APIs, the attack surface has expanded from text generation to tool execution. This domain focuses on \textit{Indirect Prompt Injection} (IPI), where the agent consumes poisoned content retrieved from external sources.

Greshake et al. \cite{greshake2023not} formalized the concept of IPI, demonstrating how retrieving a compromised website could hijack an agent's conversation. Building on this, Wang et al. \cite{wang2024allies} introduced ToolCommender, which reveals how adversaries can manipulate the tool selection process to trigger unintended functions. More recently, Zhan et al. \cite{zhan2025adaptive} proposed AdaptiveAttack, a framework that combines jailbreaking heuristics with tool manipulation to adjust payloads based on system feedback. Similarly, UDora \cite{zhang2025udora} and Breaking Agents \cite{zhang2024breaking} investigate the disruption of the internal reasoning chains of agents, causing them to enter infinite loops or execute redundant operations.

However, existing studies on tool security largely assume a scenario where the attacker controls the \textit{content} (e.g., a webpage) rather than the \textit{infrastructure} (the tool server itself). Furthermore, they often overlook the rigid schema constraints imposed by modern protocols like MCP. In this work, we specifically target the \textit{supply chain} of MCP, a domain previously underexplored. Unlike general IPI studies that treat tool outputs as unstructured text, we exploit the trust relationship inherent in structured JSON responses. Our threat model is distinct in that it focuses on a "stealthy update" scenario where a trusted tool becomes malicious, requiring a payload generation strategy that strictly adheres to schema validation while carrying a hidden adversarial instruction.

\subsection{Defense Mechanisms and Benchmarks}

The defense landscape is bifurcated into \textit{context-aware} and \textit{classification-based} strategies \cite{shi2025lessons}.

Context-aware defenses attempt to neutralize attacks via prompt engineering. Techniques such as In-Context Learning (ICL) \cite{wei2023jailbreak}, Spotlighting \cite{hines2024defending}, and Sandwich Prevention \cite{learnprompting_sandwich_defense} enclose untrusted data within safety instructions to demarcate it from system commands. Classification-based defenses employ auxiliary monitors. Perplexity filtering \cite{jain2023baseline} flags high-entropy inputs typical of GCG attacks, while methods like RTBAS \cite{zhong2025rtbas} and Task Shield \cite{jia2024task} analyze information flow to detect deviations from user intent.

To evaluate these dynamics, several benchmarks have been established. BIPIA \cite{yi2025benchmarking} and InjecAgent \cite{zhan2024injecagent} provide datasets for IPI, while Agent Security Bench (ASB) \cite{zhang2024agent} and AgentDojo \cite{debenedetti2024agentdojo} offer dynamic environments for multi-turn red teaming. Our evaluation utilizes the InjecAgent methodology but exposes a critical gap in current defenses. We demonstrate that context-aware defenses like Sandwiching fail against our multi-strategy injection, and classification-based defenses like Perplexity Filtering are ineffective against our linguistically coherent payloads. By incorporating a \textit{Defense-Aware} mechanism directly into the optimization loop, our work represents the first adaptive framework designed to systematically evade these specific countermeasures in an agentic context.

\section{Discussion}
\label{sec:discussion}

\subsection{The Reality of Probabilistic Risk at Scale}
A critical aspect of evaluating the threat posed by TIP is interpreting the significance of the Attack Success Rate (ASR). While our method achieves near-perfect success in many configurations, certain robust defense settings, such as the \texttt{ShipManager} tool under Finetuned Detection, exhibit reduced ASRs (e.g., 68.2\%). In traditional software security, a vulnerability that only triggers 68\% of the time might be considered unreliable. However, in the context of Large Language Model agents, we contend that \textit{any} non-zero ASR represents a critical vulnerability.

This argument is grounded in the scale of deployment. In real-world applications, MCP servers are designed to serve millions of users with high frequency. Risk must therefore be calculated as a function of exposure volume. An attack with a seemingly low 1\% success rate, when integrated into a popular tool invoked one million times daily, translates to 10,000 successful breaches. Unlike traditional buffer overflows which may crash a system if they fail, a failed LLM injection typically results in a benign response, allowing the attacker to remain undetected while the dice are rolled again. This \textit{stealthy persistence} grants the attacker an asymmetric advantage: they only need to succeed once to exfiltrate credentials or phishing data, whereas the defender must succeed every time.

Furthermore, the existence of these injection paths proves that the underlying trust model of MCP is fragile. The protocol assumes that tool responses are semantic data, not executable instructions. Our results demonstrate that modern LLMs, regardless of alignment training, fundamentally struggle to distinguish between the two when presented with coherent, context-aware payloads.

\subsection{Limitations and Future Works}
While TIP demonstrates state-of-the-art performance, we acknowledge certain constraints inherent to the black-box adversarial setting and outline how they pave the way for future research.

\noindent\textbf{Optimization Latency vs. Real-Time Attacks.} 
The generation of adversarial payloads via black-box search requires iterative querying of the victim model. While TIP significantly reduces the query budget by an order of magnitude compared to baselines like TAP (reducing queries from $\sim$2500 to $\sim$100), it still requires a setup phase to "train" the payload. We addressed this in our current work by designing TIP to generate "universal" payloads that are transferable. Once trained offline, these payloads can be deployed instantly during the attack phase without further optimization. Future work could focus on \textit{one-shot} generation techniques, potentially leveraging distilled attacker models that can predict effective suffixes without iterative searching.

\noindent\textbf{Single-Turn vs. Multi-Turn Persistence.}
Our current evaluation focuses on immediate injection, e.g., hijacking the agent within the very next turn after the tool response. This is the most critical vector for immediate fraud and data theft. However, in highly complex agent workflows, an attacker might aim to plant a dormant instruction that only triggers after several subsequent user interactions. 
TIP currently employs the \textit{Implicit Induction} strategy to blend into the context, which naturally aids retention. Future research should explore long-term memory injection, evaluating how effectively adversarial instructions persist in the agent's context window or external memory banks over extended sessions.

\noindent\textbf{The Defensive Arms Race.}
Our results show that static defenses and current classification-based defenses are insufficient. However, we anticipate that future defenses may employ more aggressive sanitization, such as re-paraphrasing all tool outputs via a separate, trusted LLM. We proactively designed TIP with a \textit{Defense-Aware} feedback loop, allowing it to adapt to such sanitization attempts. Future work in this domain will likely evolve into a dynamic game-theoretic struggle, where attackers and defenders continuously adapt. We envision extending TIP into a continuous learning framework that monitors defense updates in real-time and evolves its payload strategies accordingly.

\section{Conclusion} \label{sec:conclusion}

In this paper, we propose \textbf{TIP}, a black box attack framework designed to expose the fragility of trust in tool augmented Large Language Models. TIP generates stealthy and semantically coherent malicious JSON key value pairs within tool responses. By incorporating a coarse to fine strategy modeling framework and a path aware feedback mechanism, TIP demonstrates high attack effectiveness across various LLM based agent systems while maintaining linguistic fluency and contextual alignment. We validate the performance of TIP through extensive experiments. The results demonstrate that the method achieves near perfect attack success rates in settings without active defenses and remains robust under multiple sophisticated defense mechanisms. Moreover, TIP exhibits strong transferability across different models and tools, which establishes it as a practical threat in real world deployments.

Crucially, we highlight the emerging risk of such attacks specifically within the context of the Model Context Protocol (MCP). Our findings show that attackers can deploy malicious tool servers as MCP compliant services and distribute them via unofficial auto installers. Once invoked by the user, these tools silently trigger harmful behaviors in the agent, which poses a serious security challenge to the ecosystem. Our study reveals critical vulnerabilities in current tool augmented LLM systems when facing structured attacks like TIP. It emphasizes the urgent need for more robust defense mechanisms, such as response integrity verification and runtime anomaly detection. We hope this work inspires further research into securing tool based agent systems and contributes to the development of safer AI architectures in the future.

\section*{Acknowledgment}
We would like to thank the EIC, the AEs, and anonymous reviewers for their valuable comments that helped improve the quality of the paper. This work was suppported by the National Key Research and Development Program (2024YFF0618800) and the National Natural Science Foundation of China (62402114). Xudong Pan and Min Yang are the corresponding
authors. We disclose that the writing of this paper is polished using OpenAI GPT-4o. The authors have carefully proofread to make sure the content faithfully reflects the authors' original manuscript.

\bibliographystyle{IEEEtran}  
\bibliography{references}     
\end{document}